\documentclass[aps,prd,
comment,
superscriptaddress,groupedaddress,
amsmath,amssymb,floatfix,nofootinbib]
{revtex4}
\usepackage{url}
\usepackage{float}
\usepackage{graphicx}
\usepackage{bm}
\usepackage[usenames,dvipsnames]{color}
\usepackage{slashed}
\usepackage{slashed}
\begin{document}

\newcommand{\Eq}{Eq.\ }
 \newcommand{\Eqs}{Eqs.\ }
\renewcommand{\(}{\left(}
\renewcommand{\)}{\right)}
\renewcommand{\{}{\left\lbrace}
\renewcommand{\}}{\right\rbrace}
\renewcommand{\[}{\left\lbrack}
\renewcommand{\]}{\right\rbrack}
\renewcommand{\Re}[1]{\mathrm{Re}\!\{#1\}}
\renewcommand{\Im}[1]{\mathrm{Im}\!\{#1\}}
\newcommand{\dd}[1][{}]{\mathrm{d}^{#1}\!\!\;}
\newcommand{\del}{\partial}
\newcommand{\nn}{\nonumber}
\newcommand{\ie}{i.e.\,}
\newcommand{\cf}{cf.\,}
\newcommand{\refeq}[1]{Eq.~(\ref{eq:#1})}
\newcommand{\refeqs}[2]{Eqs.~(\ref{eq:#1})-(\ref{eq:#2})}
\newcommand{\reffig}[1]{Fig.~\ref{fig:#1}}
\newcommand{\refsec}[1]{Section \ref{sec:#1}}
\newcommand{\reftab}[1]{Table \ref{tab:#1}}
\newcommand{\order}[1]{\mathcal{O}\({#1}\)}
\newcommand{\fv}[1]{\left(\begin{array}{c}#1\end{array}\right)}%

\def\tcb#1{\textcolor{blue}{#1}}
\def\tcr#1{\textcolor{red}{#1}}
\def\tcg#1{\textcolor{green}{#1}}
\def\tcc#1{\textcolor{cyan}{#1}}
\def\tcv#1{\textcolor{violet}{#1}}
\def\tcm#1{\textcolor{magenta}{#1}}
\def\tcpn#1{\textcolor{pink}{#1}}
\def\tcpr#1{\textcolor{purple}{#1}}
\definecolor{schrift}{RGB}{120,0,0}

\def \azeL{{H_0^L}}
\def \azeR{{H_0^R}}
\def \apaL{{H_\parallel^L}}
\def \apaR{{H_\parallel^R}}
\def \apeL{{H_\perp^L}}
\def \apeR{{H_\perp^R}}

\newcommand{\alphas}{\alpha_\mathrm{s}}
\newcommand{\alphae}{\alpha_\mathrm{e}}
\newcommand{\gfermi}{G_\mathrm{F}}
\newcommand{\GeV}{\,\mathrm{GeV}}
\newcommand{\MeV}{\,\mathrm{MeV}}
\newcommand{\amp}[1]{\mathcal{A}\left({#1}\right)}
\newcommand{\wilson}[2][{}]{\mathcal{C}_{#2}^{\mathrm{#1}}}
\newcommand{\bra}[1]{\left\langle{#1}\right\vert}
\newcommand{\ket}[1]{\left\vert{#1}\right\rangle}

%

\let\a=\alpha      \let\b=\beta       \let\c=\chi        \let\d=\delta
\let\e=\varepsilon \let\f=\varphi     \let\g=\gamma      \let\h=\eta
\let\k=\kappa      \let\l=\lambda     \let\m=\mu
\let\o=\omega      \let\r=\varrho     \let\s=\sigma
\let\t=\tau        \let\th=\vartheta  \let\y=\upsilon    \let\x=\xi
\let\z=\zeta       \let\io=\iota      \let\vp=\varpi     \let\ro=\rho
\let\ph=\phi       \let\ep=\epsilon   \let\te=\theta
\let\n=\nu
\let\D=\Delta   \let\F=\Phi    \let\G=\Gamma  \let\L=\Lambda
\let\O=\Omega   \let\P=\Pi     \let\Ps=\Psi   \let\Si=\Sigma
\let\Th=\Theta  \let\X=\Xi     \let\Y=\Upsilon
%


\def\cA{{\cal A}}                \def\cB{{\cal B}}
\def\cC{{\cal C}}                \def\cD{{\cal D}}
\def\cE{{\cal E}}                \def\cF{{\cal F}}
\def\cG{{\cal G}}                \def\cH{{\cal H}}
\def\cI{{\cal I}}                \def\cJ{{\cal J}}
\def\cK{{\cal K}}                \def\cL{{\cal L}}
\def\cM{{\cal M}}                \def\cN{{\cal N}}
\def\cO{{\cal O}}                \def\cP{{\cal P}}
\def\cQ{{\cal Q}}                \def\cR{{\cal R}}
\def\cS{{\cal S}}                \def\cT{{\cal T}}
\def\cU{{\cal U}}                \def\cV{{\cal V}}
\def\cW{{\cal W}}                \def\cX{{\cal X}}
\def\cY{{\cal Y}}                \def\cZ{{\cal Z}}


\def\be{\begin{equation}}
\def\ee{\end{equation}}
\def\bea{\begin{eqnarray}}
\def\eea{\end{eqnarray}}
\def\bm{\begin{matrix}}
\def\em{\end{matrix}}
\def\bpm{\begin{pmatrix}}
    \def\epm{\end{pmatrix}}

{\newcommand{\lsim}{\mbox{\raisebox{-.6ex}{~$\stackrel{<}{\sim}$~}}}
{\newcommand{\gsim}{\mbox{\raisebox{-.6ex}{~$\stackrel{>}{\sim}$~}}}
\def\mpl{M_{\rm {Pl}}}
\def\gev{{\rm \,Ge\kern-0.125em V}}
\def\tev{{\rm \,Te\kern-0.125em V}}
\def\mev{{\rm \,Me\kern-0.125em V}}
\def\ev{\,{\rm eV}}

\title{\boldmath Scrutinizing $R$-parity violating interactions in light of $R_{K^{(\ast)}}$ data}
\author{Diganta Das}
\email{diganta@prl.res.in}
\affiliation{Theoretical Physics Division, Physical Research Laboratory, Navrangpura, Ahmedabad 380 009, India}
\author{Chandan Hati}
\email{chandan@prl.res.in}
\affiliation{Theoretical Physics Division, Physical Research Laboratory, Navrangpura, Ahmedabad 380 009, India}
\author{Girish Kumar}
\email{girishk@prl.res.in}
\affiliation{Theoretical Physics Division, Physical Research Laboratory, Navrangpura, Ahmedabad 380 009, India}
\author{Namit Mahajan}
\email{nmahajan@prl.res.in}
\affiliation{Theoretical Physics Division, Physical Research Laboratory, Navrangpura, Ahmedabad 380 009, India}

\begin{abstract}
The LHCb has measured the ratios of $B\to K^\ast\mu^+\mu^-$ to $B\to K^\ast e^+ e^-$ branching fractions in two dilepton invariant mass squared bins, which deviate from the Standard Model predictions by approximately $2.5\sigma$. These new measurements strengthen the hint of lepton flavor universality breaking which was observed earlier in $B\to K\ell^+\ell^-$ decays. In this work we explore the possibility of explaining these anomalies within the framework of $R$-parity violating interactions. In this framework, $b\to s\ell^+\ell^-$ transitions are generated through tree and one loop diagrams involving exchange of down-type right-handed squarks, up-type left-handed squarks and left-handed sneutrinos. We find that the tree level contributions are not enough to explain the anomalies, but at one loop, simultaneous explanation of the deviations in $B\to K^\ast\ell^+\ell^-$ and $B\to K\ell^+\ell^-$ is feasible for a parameter space of the Yukawa couplings that is consistent with the bounds coming from $B\to K^{(\ast)}\nu\bar{\nu}$ and $D^0\to \mu^+\mu^-$ decays and $B_s-\bar{B}_s$ mixing.
\end{abstract}
\maketitle
\section{Introduction}
Precision measurements of the rare decays provide excellent probes for testing new physics (NP) beyond the Standard Model (SM) of particle physics. In the SM flavor changing neutral current transitions $b\to s \ell^+\ell^-$ arise at one-loop level and 
are suppressed by the Glashow-Iliopoulos-Maiani mechanism.  To this end we study the ratio of branching ratios of $B\to K(K^\ast)\ell\ell$ decays into di-muons
over di-electrons 
\begin{equation}
R_K = \frac{\text{Br}(B \to K \mu^+\mu^-)}{\text{Br}(B \to K e^+e^-)}\,,\quad
R_{K^*} = \frac{\text{Br}(B \to K^* \mu^+\mu^-)}{\text{Br}(B \to K^* e^+e^-)} \,.
\end{equation}
In these ratios the hadronic uncertainties cancel and therefore these observables are sensitive to lepton flavor universality (LFU) violating NP \cite{Hiller:2003js}. In 2014 the LHCb Collaboration reported the measurement of $R_K$ in the dilepton invariant mass squared bin $q^{2}\in [1,6]~\text{GeV}^{2}$ to be \cite{Aaij:2014ora}
\begin{equation}\label{eq:LHCbRK}
R_{K}^{\rm{LHCb}}=0.745\pm_{0.074}^{0.090}\pm 0.036,
\end{equation}
corresponding to a $2.6 \sigma$ deviation from the SM prediction of $R_{K}^{\rm{SM}}=1.00\pm 0.01$ \cite{Bordone:2016gaq,Bobeth:2007dw}. Very recently, the LHCb Collaboration presented their first results for $R_{K^*}$ \cite{LHCb2017}
\begin{equation}\label{eq:LHCbRKst}
R_{K^*[0.045, 1.1]} = 0.66 ^{+0.11}_{-0.07} \pm 0.03\,,\quad
R_{K^* [1.1, 6]} = 0.69 ^{+0.11}_{-0.07} \pm 0.05\,,
\end{equation}
where the subscript indicates the dilepton invariant mass squared bin in $\text{GeV}^{2}$. These values correspond to $2.4\sigma$ and $2.5 \sigma$ deviations from the SM values $R_{K^*[0.045, 1.1]}^{\rm{SM}} \sim 0.93$ and $R_{K^* [1.1, 6]}^{\rm{SM}} \sim 0.99$, respectively. Combination of these results shows significant deviation from the SM which strongly hints to LFU breaking NP.

To address these anomalies we consider the low energy effective Hamiltonian for the $b\to s\ell\ell$ transition 
\begin{equation}\label{eq:Heff}
\mathcal{H}_{\rm eff} = -\frac{4G_F}{\sqrt{2}}\frac{\alpha_e}{4\pi}V_{tb}V_{ts}^\ast \sum_{i=9,10} (C_i \mathcal{O}_i + C_i^{\prime} \mathcal{O}_i^{\prime} )\, ,
\end{equation}
where the four-fermion operators are defined as
\begin{equation}
\mathcal{O}_{9(10)} = (\bar{s}\gamma^\mu P_L b)(\bar{\ell}\gamma_\mu(\gamma_5)\ell)\,, \quad \mathcal{O}_{9(10)}^{\prime} = (\bar{s}\gamma^\mu P_R b)(\bar{\ell}\gamma_\mu(\gamma_5)\ell)\, ,
\end{equation}
with $P_{L,R}=(1\mp\gamma_5)/2$ as the chiral projectors. The model-independent approach to address these tensions is to modify the Wilson coefficients $C_i=C_i^{\rm SM}+\delta C_i$, where $C_9^{\rm SM}=-C_{10}^{\rm SM}\sim 4.2$ for all the charged leptons. $\delta C_i$ and the Wilson coefficients of the chirality flipped operators $C_{9,10}^\prime$ can appear in different NP extensions and can be lepton flavor dependent. To obtain $R_K<1$ and $R_{K^\ast}<1$, as suggested by the data, one can consider NP contributions in the Wilson coefficients $(\delta C_i,C_i^\prime)$ for electrons and the muons such that either the $B\to K^\ast\mu^+\mu^-$ rate is reduced or the $B\to K^\ast e^+e^-$ is enhanced or both. However, data on $B \to K^{(\ast)} e^+e^-$ seems to be consistent with the SM predictions. Therefore, we work in a scenario where the $B\to K^{(\ast)}\mu^+\mu^-$ rates are reduced by NP contributions while the $B\to K^{(\ast)}e^+e^-$ rates are SM like. Introducing lepton superscripts in the Wilson coefficients such solutions for $(\delta C_i, C_i^\prime)$ look like
\begin{eqnarray}
\delta C_i^\mu &=& C_i^{\rm SM}-C_i^\mu \ne 0\, , \quad C_i^{\prime\mu} \ne 0\, ,\\
\delta C_i^e &=& C_i^{\rm SM}-C_i^e = 0\, , \quad C_i^{\prime e} = 0\, .
\end{eqnarray}
Following the announcement of the LHCb results \cite{LHCb2017}, several NP models have been considered in   Refs.~\cite{Capdevila,Altmannshofer,DAmico,Hiller,Geng,Ciuchini,Celis,Becirevic,Cai,Kamenik,Sala,DiChiara,Ghosh,Alok1,Alok2,Alonso,Wang,Greljo,Bonilla,Feruglio,Ellis:2017nrp,Crivellin:2017zlb,Bishara:2017pje,Alonso:2017uky,Tang,Datta:2017ezo,Hurth:2017hxg} to explain both $R_K$ and $R_{K^\ast}$ anomalies, where the above type solutions have also been considered. Note that the effective Hamiltonian (\ref{eq:Heff}) in general comprises of (pseudo)scalar and tensor operators but they are unable to explain the LHCb data \cite{Hiller:2014yaa}.

In this work we explore the possibility to explain $R_{K^{(\ast)}}$ anomalies in R-parity violating (RPV) interactions. RPV interactions have been studied previously in  Refs.~\cite{Deshpand:2016cpw, Biswas:2014gga} to accommodate $R_K$ data. In Ref.~\cite{ Biswas:2014gga} the authors assume a scenario where a tree-level exchange of left-handed up type squark  generates enhanced $b\to s e^+e^-$ rate to obtain $R_K <1$. But we note that this scenario is unable to produce both $R_K <1$ and $R_{K^{\ast}} <1 $ simultaneously. On the other hand, in Ref.~\cite{Deshpand:2016cpw} the authors studied the possibility of explaining $R_K$ anomaly within the context of RPV via one-loop contribution involving $\tilde{d}_R$. However, the authors in  Ref.~\cite{Deshpand:2016cpw} note that the severe constraints from $B\to K^{(\ast)}\nu\bar{\nu}$ make it difficult for a viable explanation of $R_K$ in this scenario. We note that there are also left-handed up-type squarks $\tilde{u}_L$ and sneutrinos $\tilde{\nu}_L$ in this model which can give additional one-loop contributions to $b\to s\ell^+\ell^-$ transition. 
We take into account their contributions, and in addition to revisiting the $R_{K}$ anomaly, we show that one can find a parameter space for the Yukawa couplings that simultaneously explain the $R_{K^{\ast}}$ anomalies. We find that this parameter space is compatible with the upper bounds on $B\to K^{(\ast)}\nu\bar{\nu}$ branching ratios. We also briefly discuss the latest experimental results on other rare $B$ and $D$ decays.

The outline of the rest of the paper is as follows. In section \ref{sec:RPV} we briefly discuss the $R$-parity violating interactions relevant for $b\rightarrow s \mu^{+}\mu^{-}$. In section \ref{sec:bsll} we discuss the one-loop contributions to $b\rightarrow s \mu^{+}\mu^{-}$ and the relevant constraints from latest experimental data for $B_{s}-\bar{B}_{s}$ mixing amplitude, and $\bar{B}\rightarrow K^{(*)} \nu \bar{\nu}$ and $D^{0}\rightarrow \mu^{+}\mu^{-}$ decays. In section \ref{sec:concl} we summarize our results and conclude.

\section{$R$-parity violating interactions \label{sec:RPV}}
In MSSM, the relevant $R$-parity violating interactions are generated through the superpotential given by \cite{Barbier:2004ez}
\begin{eqnarray}
W_{\textrm{RPV}} &=& \mu_i L_i H_u + {1\over 2} \lambda_{ijk}L_i L_j E^c_k + \lambda^\prime_{ijk}L_i Q_j D^c_k + {1\over 2} \lambda{''}_{ijk}U^c_iD^c_jD^c_k\;.
\end{eqnarray}
Here $Q_{j}$ represents the $SU(2)_L$ quark isodoublet superfield while $U^{c}_{i}$ and $D^{c}_{j}$ represent right-handed up type and down type quark isosinglet superfields respectively. $L_{i}$, $E^{c}_{i}$ denote $SU(2)_L$ lepton isodoublet and isosinglet superfields respectively. $H_{u}$ is the up type Higgs superfield that gives masses to the up type quarks. The trilinear terms contain only dimensionless parameters, while the bilinear term contains dimensionful coupling. To ensure proton stability we will assume the $\lambda''$ coupling to be zero.

Since the processes of our interest involve both leptons and quarks, we will consider the $\lambda'$ interaction term as the source of NP in this work. The interactions induced by this term at the tree and one-loop can contribute to $b\rightarrow sll$. The relevant interaction terms in the Lagrangian can be obtained by expanding the superpotential term involving $\lambda'$ in terms of fermions and sfermions as
\begin{eqnarray}{\label{rpv:lag}}
\mathcal{L} = \lambda^{\prime}_{ijk}\left ( \tilde \nu^i_L \bar d^k_R d^j_L + \tilde d^j_L \bar d^k_R\nu^i_L + \tilde d^{k*}_R \bar \nu^{ci}_L d^j_L- \tilde l^i_L \bar d^k_R u^j_L - \tilde u^j_L \bar d^k_R l^i_L - \tilde d^{k*}_R \bar l^{ci}_L u^j_L\right )\;,
\end{eqnarray}
where the sfermions are denoted by tildes, and ``c'' denotes charge conjugated fields.

\section{$b\rightarrow s\ell^+\ell^-$ in $R$-parity violating interactions \label{sec:bsll}}
One can obtain a potential tree level contribution to $b\rightarrow s\ell^+\ell^-$ via the interaction terms given in Eq. (\ref{rpv:lag}). Integrating out $\tilde{u}_{L}$, one obtains the following four fermion operator at the tree level
\begin{eqnarray}
{\cal L}_{eff} =
-{\lambda'_{ijk}\lambda^{'*}_{i'jk'}\over 2m^2_{\tilde u^j_L}}
\bar \ell^{i'}_L \gamma^\mu \ell^i_L \bar d^k_R \gamma_\mu
d^{k'}_R \label{rpv:tree}\, ,
\end{eqnarray}
where $m_{\tilde u^j_L}$ is the mass of $\tilde {u}^j_L$. For $k=2$ and $k'=3$, the  operator $(\bar s_R \gamma_\mu b_R)(\bar \ell_L \gamma^\mu \ell_L) $  contributes to $b \rightarrow s\mu^+\mu^-$. Comparing Eq. (\ref{rpv:tree}) with the $b\rightarrow s \ell^+\ell^-$ effective Hamiltonian given in Eq.~\eqref{eq:Heff} we find the Wilson coefficients $C_{9}^{\prime\ell}$ and $C_{10}^{\prime\ell}$ corresponding to the operators $(\bar s_R \gamma_\mu b_R)(\bar \ell \gamma^\mu \ell) $ and $(\bar s_R \gamma_\mu b_R)(\bar \ell \gamma^\mu\gamma_5 \ell) $ respectively to be 
\begin{equation}
\label{rpv:wilson}
C_{10}^{\prime\ell}=-C_{9}^{\prime\ell}=\frac{\lambda_{\ell j2}'\lambda_{\ell j3}'^{*}}{V_{tb}V_{ts}^{*}}\frac{\pi}{\alpha_{e}}\frac{\sqrt{2}}{4m_{\tilde{u}_L^j}^{2}G_{F}};~~~~\ell=e,\mu.
\end{equation}
We observe that for $i=i^\prime =2$ the solution $C_{10}^{\prime\mu}=-C_{9}^{\prime\mu}$ is not able to generate $R_K<1$ and $R_{K^\ast}<1$ simultaneously. So it is not possible to explain both $R_{K^{*}}$ and $R_{K^{*}}$ with tree level contributions coming from $R$-parity violating interactions. Therefore, in the rest of the paper we do not consider this contribution.

Next we will explore one-loop contributions to $b\to s\ell^+\ell^-$ to see if $R_{K}$ and $R_{K}^{*}$ anomalies can be simultaneously explained. The model-independent analysis \cite{Hiller}  shows that for simultaneous explanation of $R_{K}$ and $R_{K}^{*}$ a negative value of $C^{\rm NP \mu}_{LL}$ is favored, where $C^{\rm NP \mu}_{LL} = \delta C_9^\mu-\delta C^\mu_{10}$. If one allows only one $k$, {\it i.e} $k^\prime=k$ in Eq. (\ref{rpv:tree}) then there is no tree level contributions to $b\to s\mu^+\mu^-$ but one-loop contributions are still possible due to the exchange of $\tilde{d}_R, \tilde{u}_L$ and $\tilde{\nu}_L$ as can be seen from equation (\ref{rpv:lag}). Representative one-loop diagrams contributing to $b\rightarrow s\mu^+\mu^-$ are shown in Fig.~\ref{fig1}. The contributions coming from these box diagrams in the limit $M_W^2, m_t^2 << m_{\tilde{d}_R}^2$  give rise to 
 \begin{eqnarray}{\label{rpv:RKWC}}
C^{\rm NP \mu}_{LL}=\frac{\lambda_{23k}^{\prime}\lambda^{\prime\ast}_{23k} }{8\pi \a_e} \left(\frac{m_t}{m_{\tilde{d}^{k}_{R}}}\right)^2 - \frac{\lambda_{i3k}^{'}\lambda^{\prime\ast}_{i2k}\lambda_{2jk}^{\prime}\lambda^{\prime\ast}_{2jk}}{32\sqrt{2}\, G_F\, V_{tb}V_{ts}^{\ast} \pi \alpha_e m^2_{\tilde{d}^{k}_{R}} }
- \frac{\lambda_{i3k}^{\prime}\lambda^{\prime\ast}_{i2k}\lambda_{2jk}^{\prime}\lambda^{\prime\ast}_{2jk}}{32\sqrt{2}\, G_F\, V_{tb}V_{ts}^{\ast} \pi \alpha_e  }
\frac{\log \left(m_{\tilde{u}^{j}_{L}}^2/ m^2_{\tilde{\nu}^i_{L}}\right)}{m_{\tilde{u}^{j}_{L}}^2 - m^2_{\tilde{\nu}^i_{L}}}\, ,
\end{eqnarray}
where repeated indices $i$ and $j$ are summed over and we assume that only couplings with $k=3$ are the dominant ones.
\begin{figure}[h!]
\begin{center}
 \includegraphics[scale=0.30]{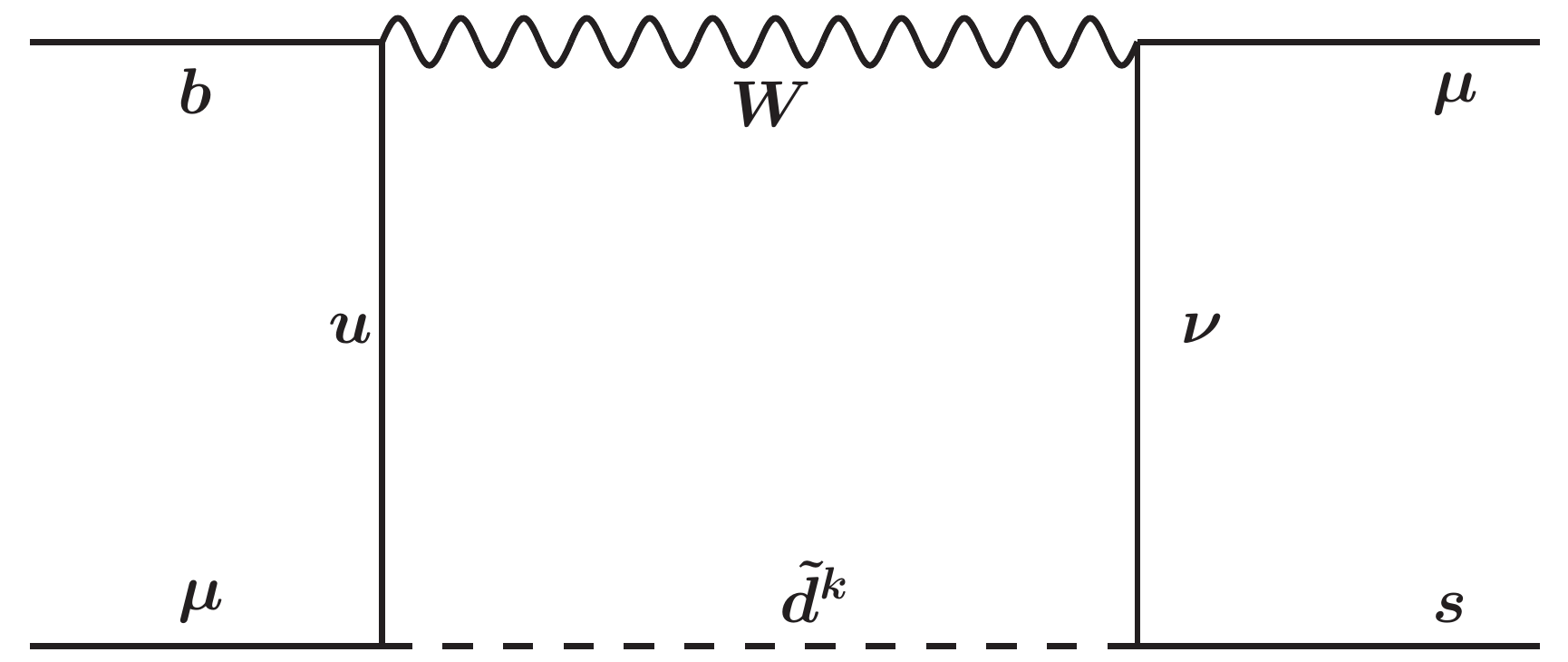}
 ~~\includegraphics[scale=0.30]{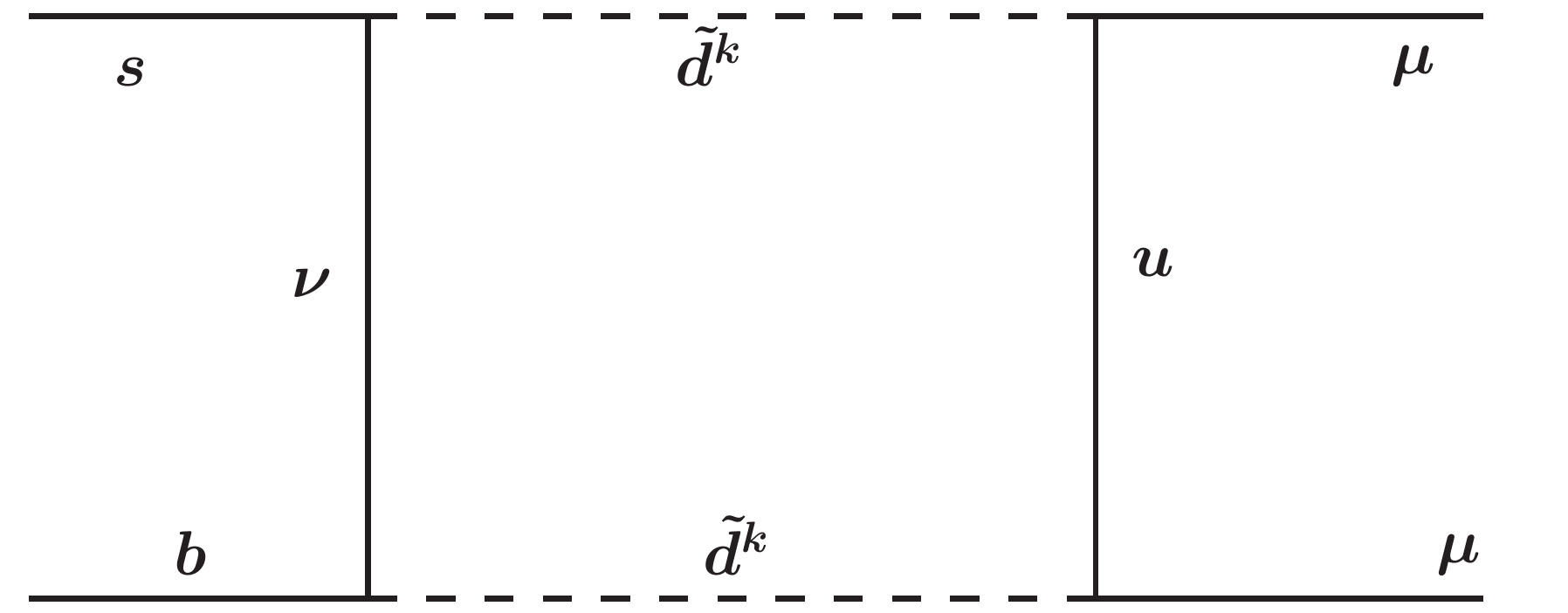}
 ~~\includegraphics[scale=0.30]{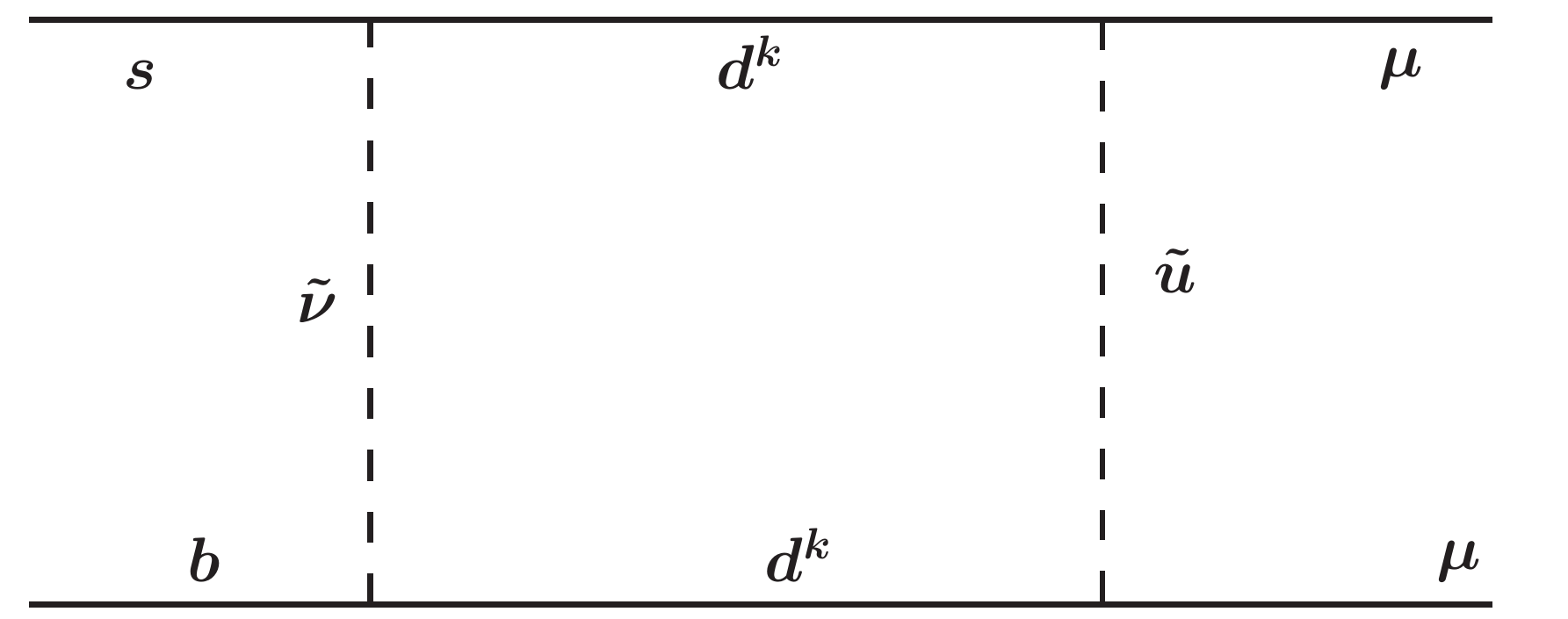}
 \caption{Representative diagrams for $b\to s\mu^+\mu^-$ transition in $R$-parity violating interactions. }
    \label{fig1}
\end{center}
 \end{figure}
Note that the first term correspond to the contribution coming from the box diagrams with a $W$ boson and $\tilde{d}^{k}_{R}$ in the loop. The second and the third terms correspond to box diagrams with two $\tilde{d}^{k}_{R}$ in the loop and its supersymmetric counterpart respectively. The first two contributions are similar to the ones in the leptoquark model discussed in Ref.~\cite{Bauer:2015knc}. We also note that the $\gamma$- and $Z$-penguin diagrams (including the supersymmetric counterparts) give vanishing contributions \cite{Lunghi:1999uk,Abada:2012cq,Krauss:2013gya,Abada:2014kba,Bauer:2015knc,Das:2016vkr}.
The last term which is the new contribution in our analysis was not included in Ref. \cite{Deshpand:2016cpw} on account of the assumption that $\tilde{u}$, $\tilde{\nu}$ are much heavier compared to $\tilde{d}_{R}$. In the absence of this contribution, the constraints from $\bar{B}\rightarrow K^{(\ast)} \nu \bar{\nu}$ proves to be too severe to explain the $b\rightarrow s \mu^{+}\mu^{-}$ induced $R_{K}$ and $R_{K^{*}}$ anomalies as noted in Ref. \cite{Deshpand:2016cpw}. Interestingly, since the the third term in equation (\ref{rpv:RKWC}) gives a negative contribution to $C_{LL}^{\rm NP\mu}$ we are able to find a parameter space for the Yukawa couplings that give $R_{K}<1$ and $R_{K^{*}}<1$ while being consistent with the latest upper bound on $\bar{B}\rightarrow K^{(*)} \nu \bar{\nu}$ branching ratios.
\begin{figure}[b]
 \includegraphics[height=3.5in]{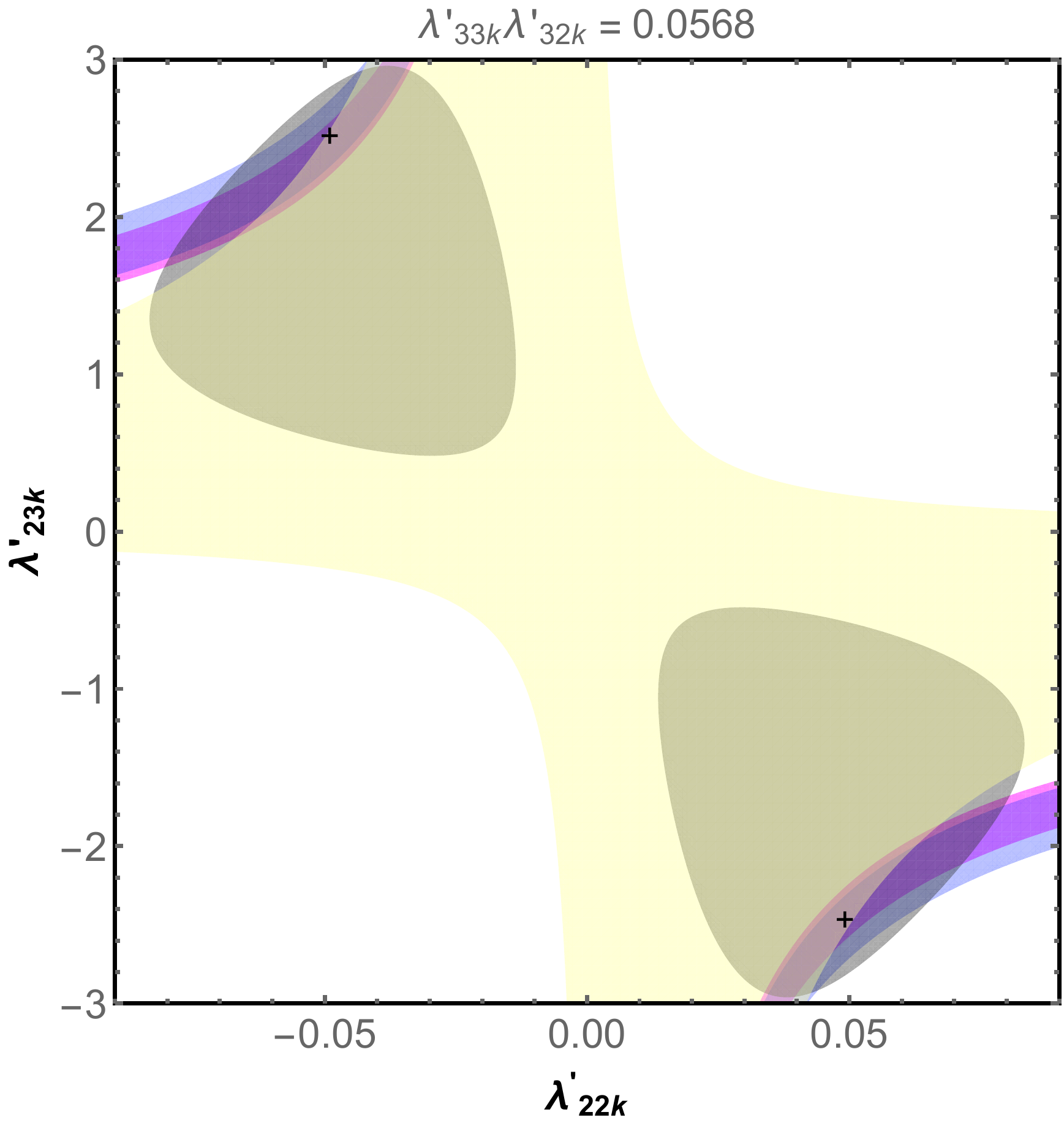} \includegraphics[height=3.5in]{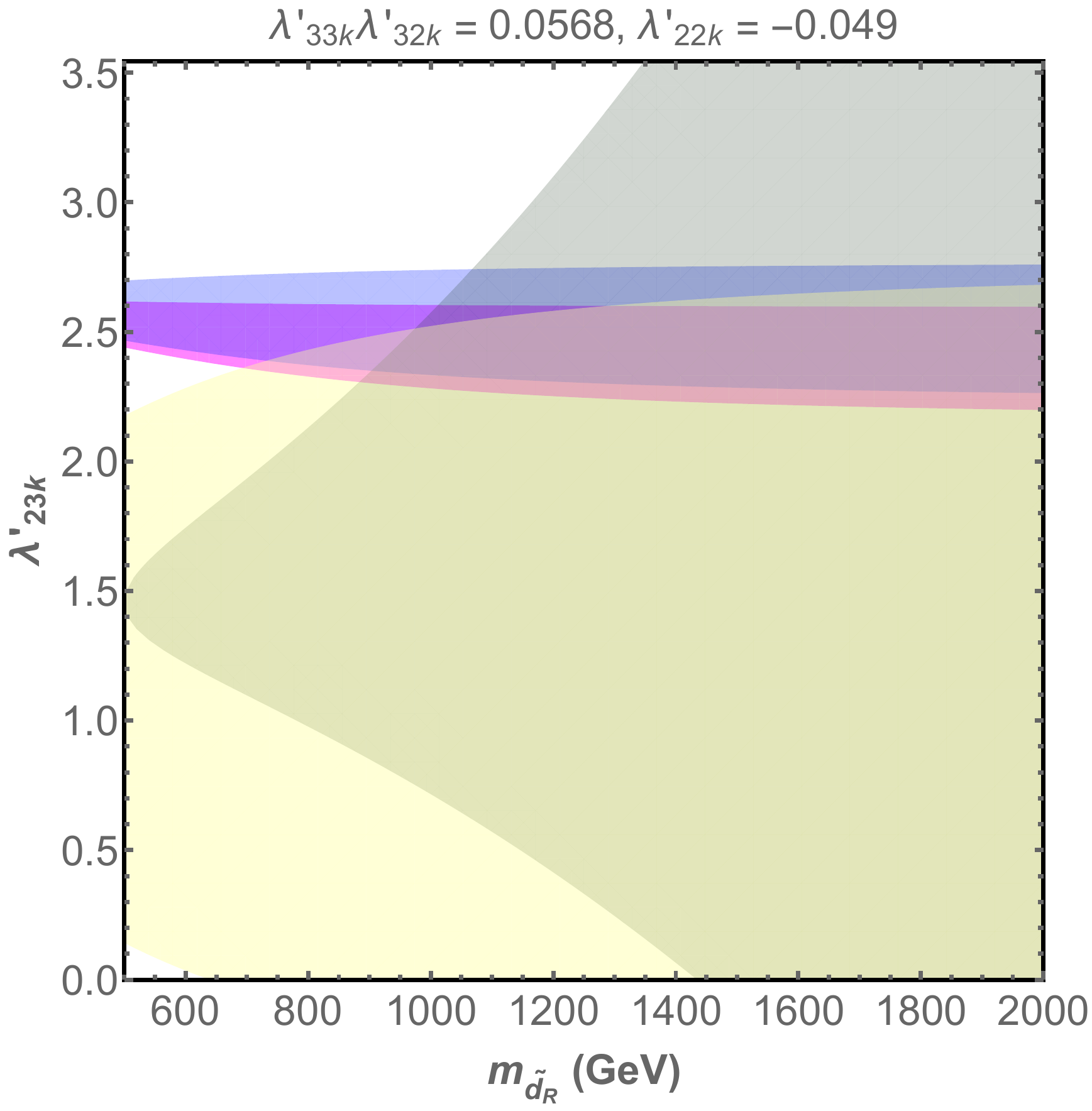}
 \caption{Plots showing the parameter space in $\lambda^{\prime}_{23k}-\lambda_{22k}^{\prime}$  plane (left) and $\lambda^{\prime}_{23k}-m_{\tilde{d}_R}$ plane (right)  explaining $R_K$ and $R_{K^{*}}$ data.  We have fixed the product of the couplings $\lambda^{\prime\ast}_{33k}\lambda_{32k}^{\prime}\sim 0.0568$ (as suggested by a $\chi^2$ analysis discussed later in the text) and take  $m_{\tilde{u}} = 1$ TeV and $m_{\tilde{\nu}}\sim 600$ GeV. For the left plot we take  $ m_{\tilde{d}} = 1.1$ TeV.   The blue (magenta) bands corresponds to the parameter space allowed by the $R_{K^{*}}$ ($R_{K}$) data.  The gray bands correspond to parameter space allowed by the $R_{B\to K(K^*) \nu\bar \nu}$ data. The light yellow band shows the parameter space compatible with  $1\sigma$ constraint from  $B_s-\bar{B}_s$ mixing. The ``cross'' mark in the left plot denotes the best fit point for the considered values of masses of $ m_{\tilde{d}}$, $m_{\tilde{u}}$, and $m_{\tilde{\nu}}$.}
    \label{fig2}
 \end{figure}

Before we study the parameter space, we discuss the constraints coming from other rare processes such as $B_s-\bar{B}_s$ mixing. RPV interations can give rise to tree level contribution to $B_s-\bar{B}_s$ due to $\tilde{\nu}$ exchange, but for specific choices of the indexes $j$ and $k$. Since we assume $k = k^\prime$, tree level contribution is absent and the leading contribution to $B_s-\bar{B}_s$ arises through one loop diagrams involving $\tilde{d}-\nu$ and $\tilde{\nu}-d$ in our scenario.
In Eq.~(\ref{rpv:RKWC}) we see that $C^{\rm NP\mu}_{ LL }$ depends on the product of couplings $\lambda_{i3k}^{\prime}\lambda^{\prime\ast}_{i2k}$ which also contributes to $B_{s}-\bar{B}_{s}$ mixing amplitude which can in turn be used to constrain these set of couplings. We follow the prescription of the UT$fit$ Collaboration \cite{Bona:2007vi} and define the ratio $C_{B_s} e^{2 i \phi_{B_s} } = \langle B_s |H^\text{full}_\text{eff} | \bar{B}_s \rangle / \langle B_s |H^\text{SM}_\text{eff} | \bar{B}_s \rangle $ which reads
\begin{equation}
 C_{B_s}e^{2i\phi_{B_s}} = 1 + \frac{m_W^2}{g^4 S_0(x_t)} \Big( \frac{1}{m^2_{\tilde{d}^k_R}} + \frac{1}{m^2_{\tilde{\nu}^i_L}} \Big) \frac{\lambda_{i3k}^\prime \lambda_{i2k}^{\prime\ast}  \lambda_{i'3k}^{\prime}\lambda_{i'2k}^{\prime\ast}}{( {V_{tb} V^{\ast}_{ts} })^2}\, .
\end{equation}
The latest UT$fit$ values of the $B_{s}-\bar{B}_{s}$ mixing parameters are $C_{B_s} = 1.070\pm0.088$ and $\phi_{B_s}=(0.054\pm 0.951)^{\circ}$ \cite{Bona:2007vi}. To be conservative, we take the upper limit of $C_{B_s}$ and find the constraint on $\lambda^{\prime}_{i3k}\lambda^{\prime\ast}_{i2k}$ to be $\left\vert\lambda^{\prime}_{i3k}\lambda^{\prime\ast}_{i2k}\right\vert\lesssim 0.067$
for $m_{\tilde{d}^k_R}\sim 1\, \text{TeV}$ and $m_{\tilde{\nu}^i_L}\sim 0.6 \,\text{TeV}$. Now these same set of couplings also contribute to  processes $\bar{B} \to K(K^{*}) \nu\bar \nu$. The ratio $R_{\bar{B} \to K(K^{*}) \nu\bar \nu} =\Gamma_{\rm RPV}(\bar{B} \to K(K^{*}) \nu\bar \nu)/\Gamma_{\rm SM}(\bar{B} \to K(K^{*}) \nu\bar \nu)$ is given by \cite{Deshpand:2016cpw}
\begin{eqnarray}\label{btoknunu}
&&R_{\bar{B} \to K(K^{*}) \nu\bar \nu}  = \sum_{i=,e,\mu,\tau}{1\over 3}\left \vert 1 +{\chi^{RPV}_{\nu_i\bar \nu_i}\over X_0(x_t) V_{tb}V^*_{ts}} \right \vert ^2 + {1\over 3} \sum_{i\neq i'} \left \vert {\chi^{RPV}_{\nu_i\bar \nu_{i'} }\over X_0(x_t)V_{tb}V^*_{ts} }\right \vert ^2\;,
\end{eqnarray}
with
\begin{eqnarray}
&&\chi^{RPV}_{\nu_i \bar \nu_{i'}} = {\pi s^2_W \over \sqrt{2} G_F \alpha} \left ( -{\lambda^{\prime}_{i3k}\lambda^{\prime *}_{i' 2 k}
\over 2 m^2_{\tilde d^k_R}} \right) \;,\;\;\;\;X_0(x_t) = {x_t(2+x_t)\over 8(x_t - 1)} + {3x_t(x_t-2)\over 8(x_t-1)^2}\ln x_t\;,
\end{eqnarray}
and $x_t = m^2_t/m^2_W$. The  RPV couplings which can modify  the rate of $B\to K^{(\ast)}\nu\nu$ appear in the following combinations,  $\lambda_{33k}^{'}\lambda^{'\ast}_{32k}$, $\lambda_{23k}^{'}\lambda^{'\ast}_{22k}$, $\lambda_{23k}^{'}\lambda^{'\ast}_{32k}$, and $\lambda_{33k}^{'}\lambda^{'\ast}_{22k}$. The latest experimental data from Belle \cite{Grygier:2017tzo} gives $R_{B\to K(K^*) \nu\bar \nu} <3.9 (2.7)$ at 90\% confidence level. Assuming one set of the product of couplings ($i=i^\prime$ or $i\ne i^\prime$) to be non-zero, the bounds on these couplings turn out to be
\bea{\label{bsnn1}}
   0.038 \left( \frac{m_{\tilde{d}_R}}{1~\text{TeV}}\right)^{2} \gtrsim \Big( \lambda^{\prime}_{23k}\lambda^{\prime\ast}_{22k}+\lambda^{\prime}_{33k}\lambda^{\prime\ast}_{32k} \Big) \gtrsim -0.079 \left( \frac{m_{\tilde{d}_R}}{1~\text{TeV}}\right)^{2} ,{\rm ~~~~~~~if~} i=i^\prime,
   \eea
and 
\begin{equation}{\label{bsnn2}}
0.055 \left( \frac{m_{\tilde{d}_R}}{1~\text{TeV}}\right)^{2} \gtrsim\Big( \lambda^{\prime}_{33k}\lambda^{\prime\ast}_{22k}+\lambda^{\prime}_{23k}\lambda^{\prime\ast}_{32k}\Big)\gtrsim -0.055 \left( \frac{m_{\tilde{d}_R}}{1~\text{TeV}}\right)^{2}{\rm ~~~~~~~if~ i\ne i^\prime},
\end{equation}

The contribution from the box diagrams also depends on one additional set of couplings $\lambda^{\prime\ast}_{2jk}\lambda_{2jk}^{\prime}$ which is always positive in our case. For $j=2, 3$, the set of couplings $\lambda^{\prime\ast}_{22k}\lambda_{22k}^{\prime}$ and $\lambda^{\prime\ast}_{23k}\lambda_{23k}^{\prime}$  are constrained from the experimental upper bound on the branching ratio for $D^{0}\rightarrow \mu^{+}\mu^{-}$, which is given by $6.2 \times 10^{-9}$ at 90\% confidence level \cite{Olive:2016xmw}. At the quark level, $D^0\to \mu^+\mu^-$ is mediated by the transition $c\to u\mu^+\mu^-$. The short-distance effective Lagrangian for $c\to u\mu^+\mu^-$  in R-parity violating interactions is given by
\begin{eqnarray}
&&\mathcal{L}_{eff} = {1\over 2 m^2_{\tilde d^k_R}} \lambda^\prime_{2jk}\lambda^{\prime *}_{2j'k} V_{1j'}V^*_{2j} \mu_L\gamma_\mu \mu_L \bar u_L \gamma^\mu c_L\;.
\end{eqnarray}
 In terms of RPV couplings, the decay width for $D^0\to \mu^+\mu^-$ is given by \cite{Deshpand:2016cpw}
\begin{eqnarray}
\Gamma(D^0\to \mu^+\mu^-) = {1 \over 128 \pi }  \left \vert {\lambda^\prime_{2jk}\lambda^{\prime *}_{2j'k} V_{1j'}V^*_{2j}\over m^2_{\tilde d^k_R}} \right \vert ^2
f_D^2 m_D m^2_\mu  \sqrt{1-{4 m^2_\mu\over m^2_D}}\;,
\end{eqnarray}
where the $D^0$ decay constant is $f_D = 212(1)$ MeV \cite{Rosner:2015wva}. Note that in the SM, the decay rate for $D^0\to \mu^+\mu^-$ is very tiny($<10^{-10}$) and we neglect it.
Taking only $\lambda^{'}_{22k}$ to be  non zero, the upper bound on $D^0\to \mu^+\mu^-$ branching ratio gives  $\lambda^{\prime\ast}_{22k}\lambda_{22k}^{\prime}<0.3 \left({m_{\tilde{d}_R}}/{1~\text{TeV}}\right)^2$, and taking only $\lambda^{\prime\ast}_{23k}\lambda_{23k}^{\prime}$ to be non zero we get a very weak bound,  $\lambda^{\prime\ast}_{23k}\lambda_{23k}^{\prime}<10^{2} \left({m_{\tilde{d}_R}}/{1~\text{TeV}}\right)^2$ which was also noted in Ref. \cite{Deshpand:2016cpw}. In this work we  fix the combination $\lambda^{\prime}_{i3k}\lambda^{\prime\ast}_{i2k}$ from the experimental data on $R_{B\to K(K^*) \nu\bar \nu}$ and $B_{s}-\bar{B}_{s}$ mixing discussed above, and explore the parameter space in terms of the other two couplings while being compatible with the bounds coming from $D^{0}\rightarrow \mu^{+}\mu^{-}$. To this end, we must mention that in this analysis we set the $R-$parity violating couplings associated with electron modes to be vanishing in view of the fact that the electron modes are consistent with the SM. We also set $\lambda^{\prime}_{i1k}$ to be zero and therefore the constraints from the processes like $K\to \pi \nu \bar{\mu}$ and $B\to \pi \nu\bar{\nu}$ will not affect our analysis. Constraints on RPV couplings can also be obtained from partonic channels like $b\bar{b}\to \tau^+\tau^-$ and $b\bar{b}\to \mu^+\mu^-$. Using the ATLAS \cite{Aad:2015osa,Aaboud:2016cre} data on di-tau final states, constraints on $({\bf 3}, {\bf 2}, 1/6)$ leptoquark model have been studied in \cite{Faroughy:2016osc}. We note that in our model $b\bar{b}\to \tau^+\tau^- (\mu^+\mu^-)$ arise at tree level via exchange of $\tilde{u}$ which shares the same gauge charges with a leptoquark weak doublet $({\bf 3}, {\bf 2}, 1/6)$.  Following \cite{Faroughy:2016osc} we find that for $m_{\tilde{u}}=1$TeV the $\tilde{u}^j_L \bar{d}^k_R l^i_L$ couplings are allowed by the current ATLAS data.

In fig.~\ref{fig2} (left plot) we show the parameter space in $\lambda^{\prime}_{23k}-\lambda_{22k}^{\prime}$ plane that is allowed by the current $R_K$ and $R_{K^\ast, [1.1-6.0]}$ data. The constraints from $B \to K^{(\ast)} \nu\nu$ and $D^0 \to \mu^+\mu^-$ are also taken into account.  We have fixed the product of the couplings $\lambda^{\prime\ast}_{33k}\lambda_{32k}^{\prime}= 0.0568$  and have taken $m_{\tilde{d}}= 1.1$  TeV,  $m_{\tilde{\nu}}= 600$ GeV, and $m_{\tilde{u}} =1$ TeV. The blue band corresponds to the regions which can explain $R_{K^\ast, [1.1-6.0]}$ data and the magenta bands correspond to the allowed regions from $R_{K}$ data. The overlapping regions correspond to the regions allowed by both $R_K$ and $R_{K^\ast, [1.1-6.0]}$ data. The gray shaded regions are allowed by the latest experimental data on $R_{B\to K(K^*) \nu\bar \nu}$, while the light yellow region corresponds to values consistent within $1\sigma$ of UT$fit$ values on $B_s-\bar{B}_s$ mixing parameters. 

 We observe that by taking a heavier mass for $\tilde{d}_R$ and while keeping $m_{\tilde{u}}$ fixed, one can find a better parameter space allowed by the considered processes in our analysis.  This is simply due to the fact that the contributions from the first two terms in the expression of $C^{\rm NP \mu}_{LL}$ in Eq.~\eqref{rpv:RKWC} are suppressed for larger values of $m_{\tilde{d}}$.  Then the third term  in Eq.~\eqref{rpv:RKWC} drives the main contribution to $C^{\rm NP \mu}_{LL}$ which is always negative in our case. This is demonstrated in  fig. \ref{fig2} (right plot) where we  show the parameter space in $\lambda^{\prime}_{23k}-m_{\tilde{d}}$ plane.  Here we have again fixed the product of the couplings $\lambda^{\prime\ast}_{33k}\lambda_{32k}^{\prime}= 0.0568$ and $m_{\tilde{u}}= 1$ TeV and $m_{\tilde{\nu}}= 600$ GeV.  The blue bands correspond to the allowed region by $R_{K^\ast, [1.1-6.0]}$ data and the magenta bands correspond to the allowed region by $R_{K}$ data. The overlapping region show the values which can explain both $R_K$ and $R_{K^\ast, [1.1-6.0]}$ data simultaneously. The gray (light yellow) shaded regions show the parameter space allowed by the latest experimental data on  $R_{B\to K(K^*) \nu\bar \nu}$ ($B_s-\bar{B}_s$ mixing). 
Note that for the above parameter space we find the range of $R^{\rm RPV}_{K^\ast, [0.045-1.1]}$ to be [0.82-0.87] which is close to the $1\sigma$ range of LHCb measurement (\ref{eq:LHCbRKst}). Moreover, values of all the couplings are well below the naive perturbative unitarity limit $\sqrt{4\pi}$. We consider this as a very good agreement.

In order to present a more robust numerical analysis, we perform a $\chi^2$-test by defining a $\chi^2$ function as
\be \label{eq:chi func}
\chi^2 = \frac{\left(R_K^{\rm Exp}-R_K^{\rm Th}\right)^2}{\left(\Delta R_K^{\rm Exp}\right)^2} + \frac{\left(R_{K^\ast , [1.1-6.0]}^{\rm Exp}-R_{K^\ast, [1.1-6.0]}^{\rm Th}\right)^2}{\left(\Delta R_{K^\ast, [1.1-6.0]}^{\rm Exp}\right)^2} + \frac{\left(R_{K^\ast, [0.045-1.1]}^{\rm Exp}-R_{K^\ast, [0.045-1.1]}^{\rm Th}\right)^2}{\left(\Delta R_{K^\ast, [0.045-1.1]}^{\rm Exp}\right)^2},
\ee
where $R_K^{\rm Exp}$, $R_{K^\ast , [1.1-6.0]}^{\rm Exp}$, and $R_{K^\ast, [0.045-1.1]}^{\rm Exp}$ refer to the central values of the experimental measurements of observables as given in \eqref{eq:LHCbRK} and \eqref{eq:LHCbRKst}. 
 $\Delta R_i^{\rm Exp}$ denote the $1\sigma$ uncertainties in the experimental measurements of observables $R_i^{\rm Exp}$ (with systematic and statistical errors added in the quadrature), while $R_i^{\rm Th}$ are the theoretical predictions of the observable. In the SM, we find $\chi^2_{\rm SM} \simeq 19$. In the considered model, the observables $R_{K, K^{\ast}}$ are functions of four new couplings $\l_{22k,23k,32k,33k}^\prime$ and masses of $\tilde{d}_R$, $\tilde{u}_L$ and $\tilde{\nu}_L$. We minimize the $\chi^2$ function subject to the conditions that the parameter space do not violate the data on the $B_s-\bar{B}_s$ mixing parameters, $b \to s \nu \nu$ and $D \to \mu \mu$ processes as discussed earlier. We also take into account of the $ b\to c (u) \ell \nu_\ell$ data that we will discuss in the next paragraph. 
   Note that $ b\to c (u) \ell \nu_\ell$ processes are very sensitive to the coupling $\l_{33k}^\prime$. Though a larger value of $\l_{33k}^\prime$ is acceptable by $b \to s \mu \mu$ data, it will produce large branching ratios for $ b\to c (u) \ell \nu_\ell$. We will comment more on this issue after discussing numerical analysis. During the minimization we keep the masses of $\tilde{d}_R$, $\tilde{u}_L$ and $\tilde{\nu}_L$ fixed. We find that in our model, with the choices $m_{\tilde{d}} = 1.1$ TeV, $m_{\tilde{u}} = 1$ TeV and $m_{\tilde{\nu}} = 0.6$ TeV, minimum $\chi^2$ is $\chi^2 \simeq 2.65$ which corresponds to the RPV couplings $\l_{22k}^\prime = -0.05$, $\l_{23k}^\prime = 2.49$, $\l_{32k}^\prime =  0.04$, $\l_{33k}^\prime =  1.42$. 
  These values of the couplings yield $C_{LL}^{\rm NP\mu} = -1.14$ and the corresponding values of the observables read $R_K = 0.74$,  $R_{K^\ast , [1.1-6.0]} = 0.73$, and $R_{K^\ast , [0.045-1.1]} = 0.84$. This is a good consistency with the experimental data on $R_K$ and $R_{K^\ast [1.1-6]}$, while the value of $R_{K^\ast}$ data in the low $q^2$ bin $ [0.045-1.1]$ lies just outside the $1\sigma$ window of experimental mean value.
One important point to note is that, by choosing slightly higher mass for $\tilde{d}_R$ or slightly lower mass for $\tilde{u}_L$ improves the fit further and a smaller $\chi^2$ value can be achieved. For example, by taking $m_{\tilde{d}_R} =1.5$ TeV and keeping other masses same as in the previous case, we find $\chi^2 = 2.44$ which correspond to $\l_{22k}^\prime =-0.06 $, $\l_{23k}^\prime =2.61 $, $\l_{32k}^\prime = 0.06$, $\l_{33k}^\prime = 1.40$. The corresponding values for observables read $R_K = 0.70$, $R_{K^\ast, [1.1-6.0]} = 0.69$, and $R_{K^\ast, [0.045-1.1]} = 0.83$. 
As also noted in the previous paragraph, this happens because the first two terms in the expression of $C_{LL\mu}^{\rm NP}$ in Eq.~\eqref{rpv:RKWC} are suppressed by $m_{\tilde{d}_R}^2$ while the last term (always negative in our case) is independent of $m_{\tilde{d}_R}$. In particular, the suppression of the first term (which is always positive in our case) helps in obtaining overall negative value of $C_{LL\mu}^{\rm NP}$ required to explain the anomalies. The higher value for $m_{\tilde{d}_R}$ also relaxes severe constraints from $B -\bar{B}$ and  $b \to s \nu \nu $ (this is shown in the second plot in Fig~\ref{fig2}).
 
We  now would like to comment on the impact of the above parameter space on the latest $B\to D^{(\ast)}\ell\bar{\nu}$ date.  The $R$-parity violating couplings can also induce new physics contribution to the semileptonic decays induced by $b\rightarrow c (u) l\nu$ where  B-factories \cite{Lees:2012xj, Lees:2013uzd,Huschle:2015rga,Sato:2016svk,Abdesselam:2016xqt,Hirose:2016wfn} and LHCb \cite{Aaij:2015yra} have measured related lepton flavor universality ratios $R_{D^{(*)}}$
 \be
 R_{D^{(\ast)}}=\frac{{\rm{Br}}(\bar{B}\rightarrow D^{(\ast)}\tau\bar{\nu})}{{\rm{Br}}(\bar{B}\rightarrow D^{(\ast)} \ell \bar{\nu})}; ~~~~~\ell = e, \mu\, .
 \ee
 The world average of the measurements for $R_{D^{*}}$ and $R_{D}$ at present is $R_{D^\ast} = 0.310 \pm 0.015 \pm 0.008$ and $R_D = 0.403 \pm 0.040 \pm 0.024$ \cite{Amhis:2016xyh}. When combined together these values differ from the SM predictions \cite{Bernlochner:2017jka,Fajfer:2012vx,Bigi:2016mdz,Lattice:2015rga,Na:2015kha} by about $4\,\sigma$. We note that the RPV interactions given in Eq.\eqref{rpv:lag} allows for tree level contribution to $b\rightarrow c (u) \ell\nu$ transitions via the exchange of down-type right handed squarks $\tilde{d}^k_R$.  
 There exists a number of studies concerning the explanation of $R_{D^{(*)}}$ experimental data within RPV scenario \cite{Deshpande:2012rr,Deshpand:2016cpw,Altmannshofer:2017poe}. The minimal setup to explain these excesses is by invoking new physics in tau mode only and having muon and electron modes SM like. However, simultaneous explanation of LFU ratios $R_{K^{(\ast)}}$ and $R_{D^{(\ast)}}$ in RPV pose a challenge, as noted in Ref.~\cite{Deshpand:2016cpw}. In our scenario for some region of the parameter space above that is consistent with $R_{K}$ and $R_{K^\ast}$ we find that ratios $R_D$ and $R_{D^{*}}$ to be SM like. Following Ref.~\cite{Deshpand:2016cpw} to study LFU  in semileptonic B-decays one can define ratios $r(B\rightarrow D^{(\ast)}\tau \nu) = R_{D^{(\ast)}}/R_{D^{(\ast)}}^{\rm SM}$ 
\be
r(B\rightarrow D^{(\ast)}\tau \nu) = \frac{2 R_\tau(c)}{R_\mu(c) + R_e(c)}\, ,
\ee
where $R_\ell(c) ={\rm BR}(B \rightarrow D^{(\ast)}\ell \nu)/{\rm BR}(B \rightarrow D^{(\ast)}\ell \nu)_{\rm SM} ~(\ell=e, \mu, \tau)$. Similarly, one can define a ratio $r(B\rightarrow\tau \nu)$ related to decay $B\rightarrow \ell \nu$ as
\be
r(B\rightarrow \tau \nu) = \frac{2 R_\tau(u)}{R_\mu(u) + R_e(u)}\, ,
\ee
with $R_\ell(u)$ given by $R_\ell(c) = {\rm BR}(B \rightarrow \ell \nu)/{\rm BR}(B \rightarrow \ell \nu)_{\rm SM}$. In the SM both $r(B\rightarrow D^{(\ast)}\tau \nu)$  and $r(B\rightarrow \tau \nu)$ are 1. The current experimental data showing  enhanced ratios for $R_D$ and $R_{D^{*}}$ with respect to the SM prefers  $r(B\rightarrow D^{(\ast)}\tau \nu)$ to be about $\sim 1.25$  \cite{Deshpand:2016cpw,Altmannshofer:2017poe}.  As a standard benchmark point, for a right handed sbottom of mass 1.1 TeV and taking previously obtained best fit point ($\l_{22k}^\prime=-0.05$, $\l_{23k}^\prime=2.49$, $\l_{32k}^\prime=0.04$, $\l_{33k}^\prime=1.42$) for the couplings, we find $r(B\rightarrow D^{(\ast)}\tau \nu)$ and $r(B\rightarrow \tau \nu)$ to be $\sim 1.04$. The individual decay rates for $B\rightarrow D^{(*)}\tau\nu$, $B\rightarrow D^{(*)}\mu\nu$, $B\rightarrow \mu\nu$, $B\rightarrow \tau\nu$ are also under control and are allowed to be enhanced at most by $10\%$ with respect to the SM, which is acceptable given the uncertainties in both experimental data and the SM predictions for these decay modes. We note that one can accommodate the current experimental data for $R_D$ and $R_{D^{*}}$ by taking a somewhat larger value of coupling $\l_{33k}^\prime$. However, larger  $\l_{33k}^\prime$ will also induce large enhancement in the decay rate of  $B\rightarrow \tau\nu$  which has not been seen in the experiments. Therefore a simultaneous explanation of  LFU ratios related to $b\rightarrow s \ell^+ \ell^-$ and $b\rightarrow c \ell \nu$ remains a challenge in our scenario.


\section{Conclusions \label{sec:concl}}
The recent LHCb results on $R_K$ and $R_{K^\ast}$ hint to lepton flavor universality breaking NP. In this work we have explored the possibility of addressing these anomalies in the framework of $R$-parity violating interaction. In our scenario, where we assume that NP enter only in the couplings of muons to the (axial)vector operators while the couplings of the electron remain SM like, we find that the tree level contributions to $b\to s\mu^+\mu^-$ transition are not able to simultaneously yield $R_K<1$ and $R_{K^\ast}<1$. 
Beyond the tree level, one-loop contributions to $b\to s\mu^+\mu^-$ are generated by the exchange of $\tilde{d}_R$, $\tilde{u}_L$ and $\tilde{\nu}_L$, which lead to a parameter space for the Yukawa couplings that can simultaneously accommodate $R_K$ and $R_{K^\ast,[1.1-6]}$ data while there is a good agreement between $R_{K^\ast,[0.045-1.1]}^{\rm RPV}$ and the measured value of $R_{K^\ast,[0.045-1.1]}$ by the LHCb. The parameter space is also consistent with the constraints coming from $B\to K^{(\ast)}\nu\bar{\nu}$ and $D^0\to\mu^+\mu^-$ decays and $B_s-\bar{B}_s$ mixing.

\begin{acknowledgements}
{\bf{Acknowledgments---}}The authors would like to thank N.G. Deshpande and Xiao-Gang He for many valuable and helpful communications. We also thank Anjan Joshipura for useful discussions.
\end{acknowledgements}

     \end{document}